\documentclass[reprint,prb,aps,groupedaddress,amsmath,amsfonts]{revtex4-2}
\usepackage[english]{babel}
\usepackage{graphicx}
\usepackage{hyperref}

\DeclareMathOperator{\sech}{sech}

\begin{document}

\title{Temperature dependent diamagnetic\textendash{}paramagnetic transitions in metal/semiconductor quantum rings}

\author{Neal \surname{Blackman}}
\affiliation{College of Engineering and Science, Louisiana Tech University, Ruston, LA 71272, USA}

\author{Dentcho A. \surname{Genov}}
\affiliation{College of Engineering and Science, Louisiana Tech University, Ruston, LA 71272, USA}
\email{dgenov@latech.edu}
\date{\today}

\begin{abstract}
We present theoretical studies of temperature dependent diamagnetic-paramagnetic transitions in thin quantum rings. Our studies show that the magnetic susceptibility of metal/semiconductor rings can exhibit multiple sign flips at intermediate and high temperatures depending on the number of conduction electrons in the ring ($N$) and whether or not spin effects are included. When the temperature is increased from absolute zero, the susceptibility begins to flip sign above a characteristic temperature that scales inversely with the number of electrons according to $N^{-1}$ or $N^{-1/2}$, depending on the presence of spin effects and the value of $N\,\mathrm{mod}\,4$. Analytical results are derived for the susceptibility in the low and high temperature limits, explicitly showing the spin effects on the ring Curie constant.
\end{abstract}

\maketitle

\section{Introduction}
The electronic properties of low-dimensional structures with ring geometry have been a subject of great interest for many decades, starting with the early study of aromatic ring currents in benzene-like compounds \cite{Pauling:1936ga,LondonF:1937bj} and later with persistent currents in microscopic conducting rings \cite{Hund:1938gz,Buttiker:1983wd}. Recent improvements in micro and nano-fabrication methods have renewed experimental interest in mesoscopic and quantum rings \cite{Levy:1990to, Chandrasekhar:1991wd, Mailly:1993jy, Lorke:2000tq, Kim:2016dj, VanDongPham:2019jh, BleszynskiJayich:2009uo, Bluhm:2009ig, Deblock:2002ir, Jariwala:2001kf, Fuhrer:2001to, Kleemans:2007em}, and in the past several decades numerous theoretical studies have continued to advance our understanding of these low dimensional metallic and semiconductor structures (see e.g. \cite{Buttiker:1983wd, Chakroborty:1994wv, Ghosh:2013bv, Murzaliev:2019df, Viefers:2004eh, Saminadayar:2004wx, Bouchiat:1989gd, Shanks:2011uo} and references therein). The use of quantum rings in real-world applications has also quickly advanced in relation to plasmonic devices and metamaterials \cite{Monticone:2014jk, Kante:2012vk, McEnery:2014fr}, which take advantage of the enhanced electromagnetic properties of conducting nanostructures by arranging them in precisely-controlled patterns. Effective design of these materials relies on predicting the electric and magnetic susceptibilities of the nanostructures according to their shape, size, material, and temperature. Although quantum effects on the electric properties of metallic nanoparticles have been studied extensively \cite{Genzel:1975wh,Vollmer:1995vu,Quinten:2010vd,Blackman:2018ke}, the magnetic properties of nanoscopic rings are still not fully understood. In fact, as many others have pointed out \cite{BarySoroker:2008hv, Waintal:2008hw, GomezViloria:2018dr, Machura:2010im, Maiti:2006ia, Murzaliev:2019df}, past estimates of the persistent current in conducting rings have differed from experimental measurements by orders of magnitude, and even predicting the sign of the low-field susceptibility has been problematic. Possible theoretical explanations for the discrepancies have included spin-breaking due to magnetic impurities \cite{BarySoroker:2008hv}, electron interactions \cite{Waintal:2008hw}, and experimental parameters like temperature or non-uniform probability of the number of electrons on the ring \cite{Machura:2010im}.

While earlier works have provided example calculations demonstrating the possibility of diamagnetic-paramagnetic transitions governed by temperature \cite{Weisz:1994ea, Machura:2010im}, they do not provide a systematic study of this effect taking into account the influence of the ring’s size and material. Furthermore, these prior studies entirely disregard spin-induced Zeeman splitting of the energy levels, which we find can have a profound impact even for weak fields. In this paper we perform a systematic investigation of size and temperature effects on the ring susceptibility with and without spin effects. We find that as the temperature increases from absolute zero, the sign of the susceptibility can flip either once or multiple times, depending on the number of electrons on the ring ($N$) and whether or not spin effects are included. This diamagnetic-paramagnetic transition occurs above a certain critical transition temperature $T^{*}$ which decreases with the electron number according to either $N^{-1}$ or $N^{-1/2}$ power laws.

\section{Grand-canonical approach to quantum rings\label{sec:gca}}
We now consider the case of a quantum ring consisting of a fixed number $N$ of non-interacting electrons immersed in an external magnetic field. As has been well-established by now \cite{Cheung:1988wv, Bouchiat:1989gd, Altshuler:1991un, EntinWohlman:1992up, Yip:1996ww}, it is most appropriate to use the canonical ensemble when the number of particles on the ring is fixed, but it is often a good approximation to treat the ring using a grand canonical ensemble and Fermi-Dirac statistics. Although there are known differences between the two approaches, we do not expect them to affect the qualitative trends discussed in this paper. Hence, in this work we present finite-temperature calculations in the grand canonical ensemble, and a more detailed discussion of the differences between the ensembles is reserved for a later work.

The thermodynamic properties of the system can be calculated using the grand canonical potential $\Omega $, which for fermions can be written as
\begin{equation}
\label{eq:grand_potential}
\Omega \left(\mu ,\kappa ,T\right)=k_{B}T\sum _{m,\sigma _{z}}g_{m,{\sigma _{z}}}\ln \left[1-f_{m,{\sigma _{z}}}\left(\mu ,\kappa ,T\right)\right]
\end{equation}
where $f_{m,{\sigma _{z}}}\left(\mu ,\kappa ,T\right)=\left(1+\mathrm{exp}\left[\left(\epsilon _{m,{\sigma _{z}}}-\mu\right)/{k_{B}T}\right]\right)^{-1}$ is the Fermi-Dirac distribution, and $\epsilon _{m,{\sigma _{z}}}$ and $g_{m,{\sigma _{z}}}$ are the energy levels of the system and their respective degeneracy factors. The chemical potential $\mu $ is obtained from the thermodynamic average of the system total electron number  
\begin{equation}
\label{eq:electron_number_sum}
N=-\frac{\partial \Omega }{\partial \mu }=\sum _{m,\sigma _{z}}g_{m,{\sigma _{z}}}f_{m,{\sigma _{z}}}\left(\mu ,\kappa ,T\right)\textrm{.}
\end{equation}
The energy levels $\epsilon _{m,{\sigma _{z}}}$ are found from the Schr\"{o}dinger\textendash{}Pauli equation $\hat{\mathcal{H}}\psi _{m,{\sigma _{z}}}=\epsilon _{m,{\sigma _{z}}}\psi _{m,{\sigma _{z}}}$. For a one-dimensional (1D) quantum ring with radius $R$ immersed in a constant magnetic field $\vec {B}=B\hat{z}$  perpendicular to the plane of the ring, the Hamiltonian reads as $\hat{\mathcal{H}}=\epsilon_1 \left(\left(-i\partial _{\phi }+\kappa \right)^{2}+g_{s}\sigma _{z}\kappa \right)$, where $\epsilon_1 =\mathrm{\hbar }^{2}/2m_{e}R^{2}$ is the zero-field energy spacing between the ground level and the next-highest level, and $\kappa =\Phi /\Phi _{0}$ is a flux parameter written in terms of the magnetic flux through the ring $\Phi =\pi R^{2}B$ and the flux quantum $\Phi _{0}=2\pi \hbar /e$. In this study we also account for the interaction of the electron’s spin with the external magnetic field (Zeeman splitting) using the effective Land\'{e} $g$-factor, $g_{s}$, and the spin operator $\sigma _{z}=\pm 1$. The eigenstates and energies of the conduction electrons then follow as 
\begin{align}
\begin{split}
\label{eq:eigenenergies}
\psi _{m,{\sigma _{z}}}&=\frac{1}{\sqrt{2\pi R}}\chi _{{\sigma _{z}}} e ^{im\phi } \\ \epsilon _{m,{\sigma _{z}}}\left(\kappa \right)&=\epsilon_1 \left(\left(m+\kappa \right)^{2}+g_{s}\sigma _{z}\kappa \right)
\end{split}
\end{align}
where $m=0,\pm 1,\cdots $ is the azimuthal quantum number and $\chi_{\sigma_z}$ is the spin part of the wavefunction. Because we consider classical free electrons, we can take $g_{s}=2$ to a high degree of accuracy, but retaining the factor as a parameter has a couple of advantages. First, it allows the model to be extended naturally to some semiconductor materials in which the effective Land\'{e} factor differs from that of free electrons. A second advantage is that in the model we can switch on or off the Zeeman splitting effect as needed. This is important in the study of persistent currents and Aharonov-Bohm rings where the magnetic field is presumed to only penetrate the interior of the ring. Thus, the two cases $g_{s}=0$ and $g_{s}=2$ represent the two extremes of spin effects turned completely on or off. 

Once the grand canonical potential is obtained we find the magnetic susceptibility of the ring which follows from thermodynamic considerations as
\begin{align}
\label{eq:susceptibility_from_potential}
\begin{aligned}
\chi &=\left.-\frac{\mu _{0}}{V}\left(\frac{\partial ^{2}\Omega }{\partial B^{2}}\right)_{T,\mu }\right| _{B=0} \\ &=\left.-\frac{\mu _{0}\mu _{B}^{2}}{4\epsilon_1 ^{2}V}\left(\frac{\partial ^{2}\Omega }{\partial \kappa ^{2}}\right)_{T,\mu }\right| _{\kappa =0}
\end{aligned}
\end{align}
Applying Eq.~(\ref{eq:susceptibility_from_potential}) to Eq.~(\ref{eq:grand_potential}) with the energy levels in Eq.~(\ref{eq:eigenenergies}) and noting that the degeneracy due to Zeeman splitting of the energy levels is $g_{m,{\sigma _{z}}}=1$ if $g_{s}\neq 0$ and $g_{m,{\sigma _{z}}}=2$ if $g_{s}=0$, we obtain
\begin{align}
	\begin{aligned}
\label{eq:susceptibility_sum}
\chi =& \left| \chi _{L}\right| \left(-1+\frac{16T_{F}}{N^{3}T}\sum _{m=-\infty }^{\infty }\left(m^{2}+\frac{g_{s}^{2}}{4}\right) \right. \\ & \hspace{6.5em} \times \left. \sech^{2} \left(\frac{8T_{F}}{N^{2}T}\left(m^{2}-\tilde{\mu }\left(0\right)\right)\right)\right)
	\end{aligned}
\end{align}
where the Langevin susceptibility due to Larmor precession is $\chi _{L}=-N\pi R^{2}\alpha ^{2}a_{B}/V=-N\chi _{1}$ where $a_{B}$ denotes the Bohr radius and $\alpha $ is the fine-structure constant. In writing Eq.~(\ref{eq:susceptibility_sum}) we have used tilde notation to indicate the dimensionless chemical potential at zero field $\tilde{\mu }\left(0\right)=\mu \left(\kappa =0\right)/\epsilon_1 $ and defined the ring 1D Fermi temperature
\begin{equation}
\label{eq:fermi_temp}
\epsilon _{F}=k_{B}T_{F}=\epsilon_1 \left(\frac{N}{4}\right)^{2}
\end{equation}
with $k_B$ Boltzmann’s constant. Note that this one-dimensional Fermi temperature follows from evaluating Eq.~(\ref{eq:electron_number_sum}) in the limit $N\rightarrow \infty $ and should not be confused with the Fermi temperature of bulk material.

\section{\label{sec:results}Ring size and temperature dependence of the susceptibility}
Here we study the characteristics of the susceptibility as a function of the temperature and the total number of conduction electrons $N$. Figure 1 shows Eq.~(\ref{eq:susceptibility_sum}) evaluated for some exemplary cases at different fixed temperatures, with and without spin effects included. The results show four distinct cases for the susceptibility depending on the value of $N\,\mathrm{mod}\,4$, which we label in Fig. 1 using the integer $j=0,1,2,\ldots $ . These four cases correspond to the possible number of paired spin-1/2 particles following the Pauli exclusion principle. This well-known $N\,\mathrm{mod}\,4$ property of one-dimensional rings is sometimes referred to as a \textit{double-parity effect} \cite{Loss:1991wy, Weisz:1994ea}. As seen in Fig. 1(a), all $N=4j+2$ cases show diamagnetic behavior for small $N$. This phenomenon is known as H\"{u}ckel’s rule in the context of aromatic chemistry \cite{Huckel:1931vw, Rickhaus:2020dr} and represents the case where all electron spins are paired. By contrast, the other set of even-numbered rings ($N=4j+4$) are paramagnetic, similar to odd-numbered rings at the chosen temperatures. In the case of a H\"{u}ckel type ring the magnetic susceptibility is found to follow the Langevin susceptibility $\chi =\chi _{L}$ until reaching $N\approx T_{F}/T$ and then decays with increasing $N$ (see Fig. 1(a)). The maximum diamagnetic response can also be estimated at $\chi _{\max }\approx -\chi _{1}T_{F}/T$. If $h$ is the physical thickness of the ring then $\chi _{1}=-\alpha ^{2}a_{B}/h$. For any physical conducting ring we have $h\geq a_{B}$, and hence the maximum diamagnetic susceptibility is given by the rather simple result $\chi _{\max }\approx -N\alpha ^{2}$. 
\begin{figure*}
\includegraphics{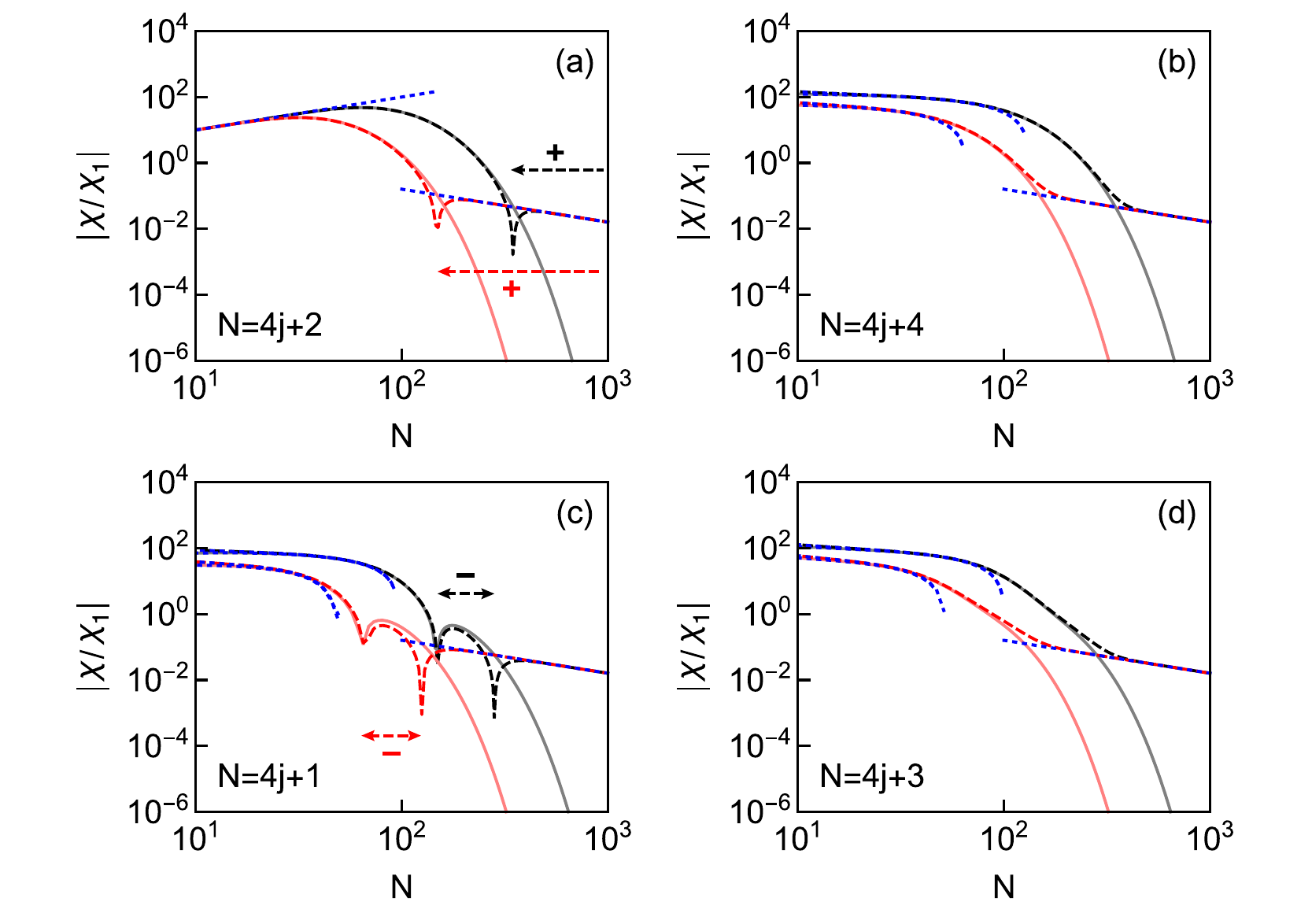}
\caption{
\label{fig:N_dependence}
The magnetic susceptibility with spin (dashed lines) and without spin (solid lines), shown at two fixed temperatures, $T/T_{F}=0.015$ (black) and $T/T_{F}=0.03$ (red). \textbf{Even number of electrons:} (a) $N=4j+2$ (H\"{u}ckel) and (b) $N=4j+4$. \textbf{Odd number of electrons:} (c) $N=4j+1$ and (d) $N=4j+3$. Arrows indicate the regions in which the susceptibility flips sign when spin effects are present. Blue dotted lines indicate the limits for $N\ll T_{F}/T$ and $N\gg T_{F}/T$ when spin is included.
}
\end{figure*}
We also see that spin has a significant impact compared to orbital effects when $N\gg T/T_{F}$. For fixed temperature and increasing $N$, the susceptibility decays exponentially if there is no spin and decays with a power law $1/N$ when spin is present. In the $N=4j+1$ and $N=4j+2$ cases, when $N\rightarrow \infty $ the sign of the susceptibility depends on whether spin effects are included. For large $N$, the $N=4j+2$ rings are diamagnetic without spin included, but when spin is included, they transition to paramagnetism above a critical size. The limiting behavior for both small and large $N$, shown as blue dotted lines in Fig. 1, follow from the low and high temperature limits derived in the sections that follow.

Finally, in Fig. 1 we also observe a strong temperature dependence with the magnetic susceptibility increasing in absolute values as the temperature is lowered. With this in mind, we now turn to the primary focus of this paper, which is to better understand the susceptibility across a broad range of temperatures. In Fig. 2 we again evaluate Eq.~(\ref{eq:susceptibility_sum}) numerically, this time keeping $N$ fixed and varying the temperature. The H\"{u}ckel-type rings in Fig. 2(a) acquire a consistent diamagnetic value $\chi=\chi _{L}$ at low temperatures, with or without spin. By contrast, the non-H\"{u}ckel rings in Fig. 2(b) have Curie-like $1/T$-dependence at low temperatures. In all four panels of Fig. 2, we find that at high temperatures the susceptibility decays exponentially without spin but decays more slowly when spin is present, acquiring a nearly constant value for $T\ll T_{F}$. The visible “kinks” in the logarithmic plots reveal that the susceptibility can flip sign at intermediate temperatures. These results demonstrate that there are three recognizable temperature regimes: low temperature, high temperature, and intermediate temperature. We further analyze each regime separately in the following three sections.
\begin{figure*}
\includegraphics{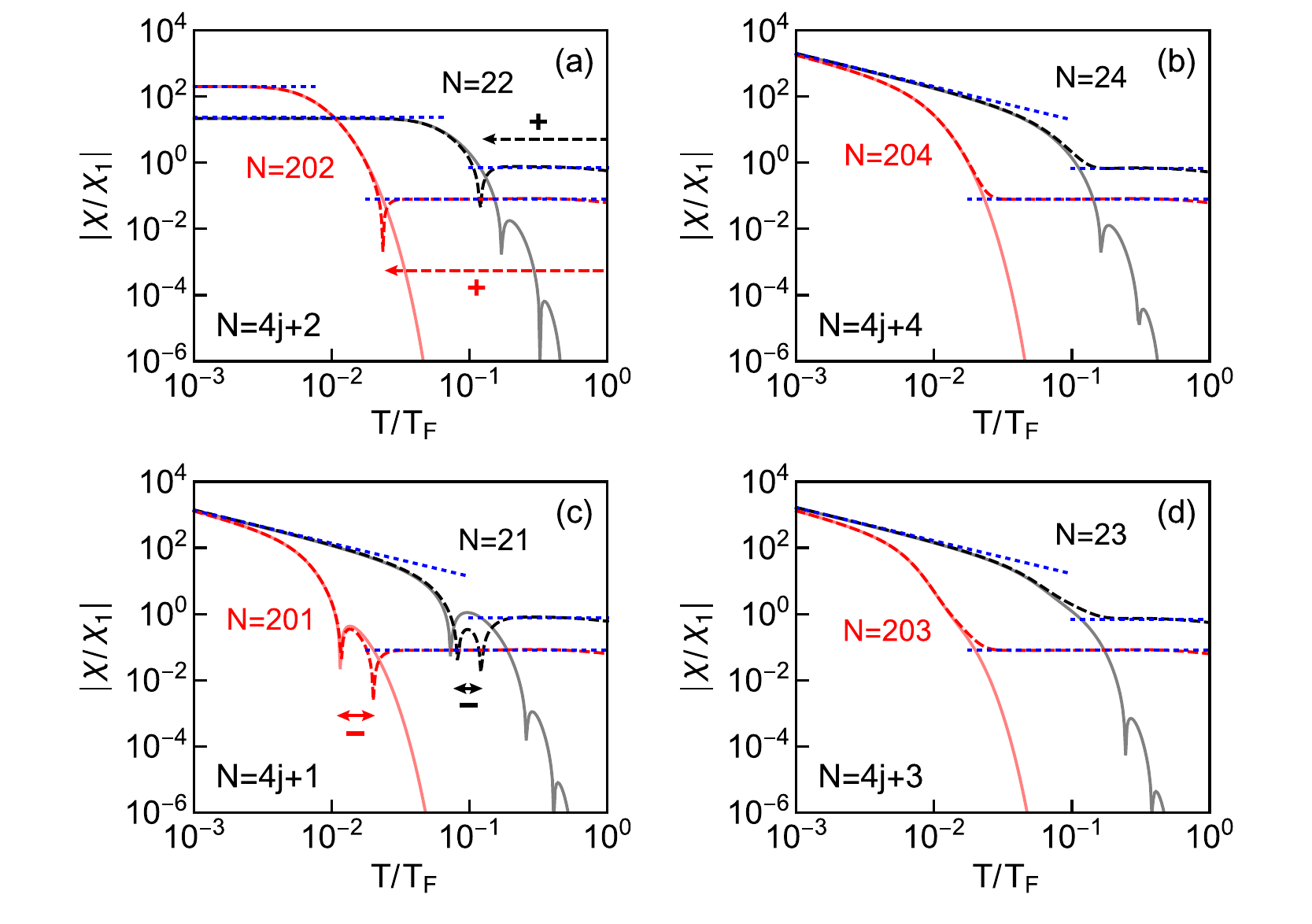}
\caption{
\label{fig:T_dependence}
The temperature dependence of the magnetic susceptibility with spin (dashed lines) and without spin (solid lines). \textbf{Even number of electrons:} (a) $N=4j+2$ (H\"{u}ckel) and (b) $N=4j+4$. \textbf{Odd number of electrons:} (c) $N=4j+1$ and (d) $N=4j+3$. Arrows indicate the regions in which the susceptibility flips sign when spin effects are included. Blue dotted lines indicate the limits for $T\gg T_{F}/N$ (when spin is included) and $T\ll T_{F}/N$ (with or without spin).
}
\end{figure*}

\begin{figure*}
\includegraphics{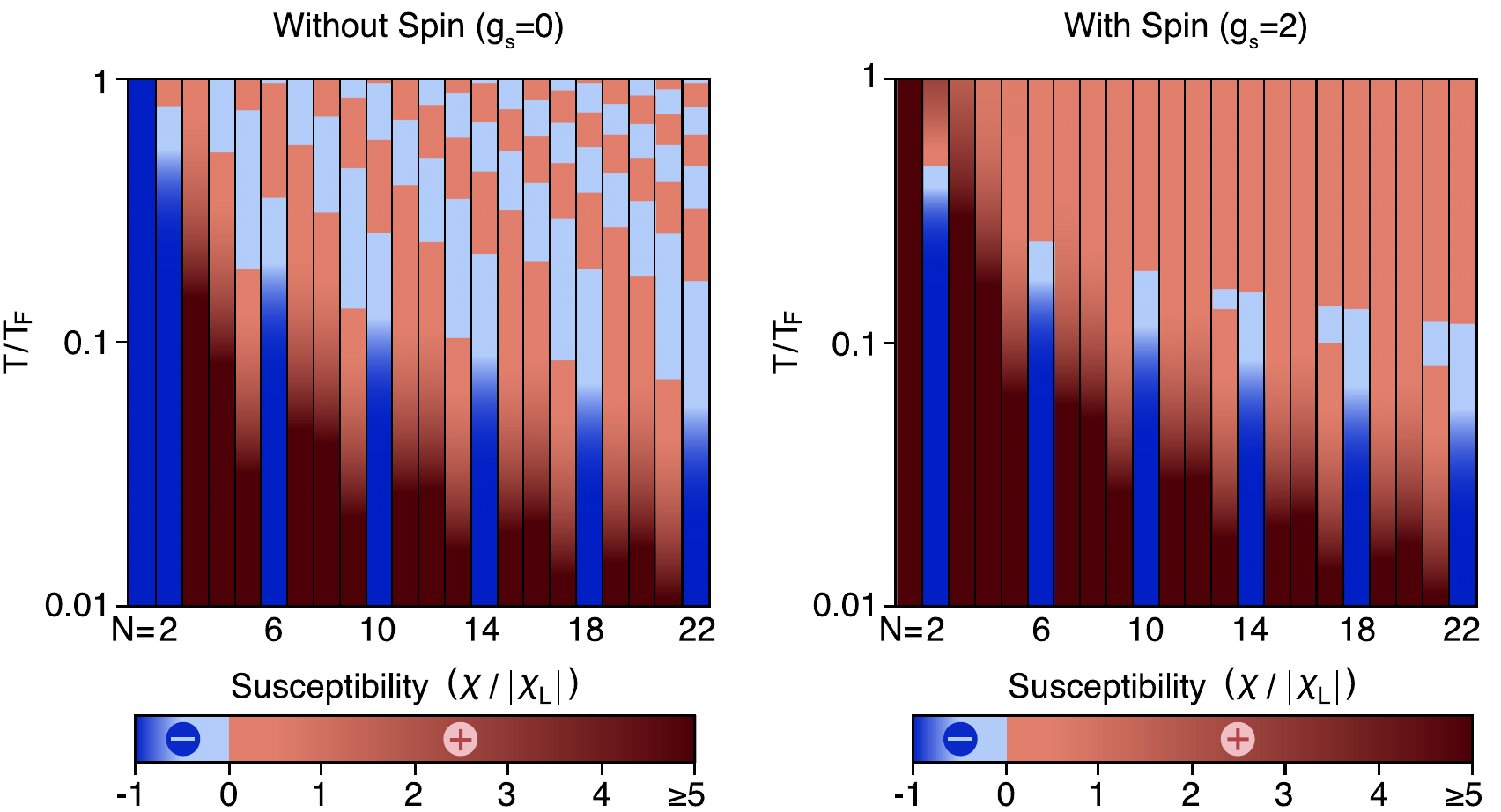}
\caption{
\label{fig:contour_plot}
The susceptibility as the temperature and number of electrons are varied, shown (left) without spin effects and (right) with spin effects. Blue regions indicate diamagnetic susceptibility and red regions are paramagnetic.
}
\end{figure*}

\vspace{-4mm}

\subsection{Low temperature limit}
When $T\ll T_{F}/N$, the Fermi-Dirac function $f_{m,{\sigma _{z}}}\left(\mu ,\kappa ,T\right)$ acts like a step function, and only the $N$ lowest energy levels are occupied. In this low-temperature limit, the chemical potential depends sensitively on $N$ and takes on four possible cases
\begin{equation}
\label{eq:low_temp_mu}
\mu \left(0\right)=\epsilon_1 \tilde{\mu }\left(0\right)=\frac{\epsilon_1 }{16}\begin{cases} \left(N-1\right)^{2}, & N=4j+1\\
\left(N^{2}+4\right), & N=4j+2\\
\left(N+1\right)^{2}, & N=4j+3\\
N^{2}, & N=4j+4
\end{cases}
\end{equation}
where again we find a double-parity effect depending on $N\,\mathrm{mod}\,4$. For $N\rightarrow \infty $ we simply have $\mu \left(0\right)=\epsilon_1 \tilde{\mu }\left(0\right)=\epsilon_1 (N/4)^{2}$, which corresponds to the Fermi temperature defined earlier in Eq.~(\ref{eq:fermi_temp}). Applying the same limit $T\ll T_{F}/N$ to Eq.~(\ref{eq:susceptibility_sum}), the summand acts like a delta function peaked at $m^{2}=\tilde{\mu }$. Thus only the $m^{2}=\tilde{\mu }\left(0\right)$ term contributes at low temperatures, where $\tilde{\mu }$ is given by Eq.~(\ref{eq:low_temp_mu}). Consequently, the susceptibility depends sensitively on the number of particles on the ring (modulo 4) and can be written compactly in the form
\begin{equation}
\label{eq:low_temp_limit}
\chi=\left| \chi _{L}\right|\left(-1+\frac{8\eta T_{F}}{N^{3}T}\left(\frac{1}{4}\left(N+\sigma _{p}\right)^{2}+g_{s}^{2}\right)\right)
\end{equation}
where $\eta =0$ if $N=4j+2$, $\eta =1$ for $N=4j+4$, and $\eta =3/4$ for odd numbers of particles. The parity factor takes the values $\sigma _{p}=-1$ for $N=4j+1$, $\sigma _{p}=+1$ for $N=4j+3$, and $\sigma _{p}=0$ when $N$ is even. From Eq.~(\ref{eq:low_temp_limit}) and in the case of a ring with all paired spins (satisfying H\"{u}ckel’s rule $N=4j+2$) we recover the characteristic Langevin diamagnetism $\chi=\chi _{L}$ as seen for low temperatures in Fig. 2(a) and small $N$ in Fig. 1(a). Note that Eq.~(\ref{eq:low_temp_limit}) is also presented as the limit for small $N$ in Fig. 1 where $N\ll T_{F}/T$. When the number of conduction electrons does not satisfy H\"{u}ckel’s rule ($\eta \neq 0$), a Curie-Weiss type of paramagnetic response is driven by the dominant $1/T$ term in Eq.~(\ref{eq:low_temp_limit}), which reduces to
\begin{equation*}
\chi=\frac{8 \eta T_{F} \left| \chi _{L}\right|}{N^{2}T}\left(\frac{N}{4}+\frac{g_{s}^{2}}{N}\right)=\frac{C}{T}
\end{equation*}
When $N\rightarrow \infty $, the effect of the spin is negligible, and we recover a Curie constant
\begin{equation}
\label{eq:curie_constant}
C_{\infty }=\frac{2\eta T_{F}\left| \chi _{L}\right| }{N}
\end{equation}
The results of Eq.~(\ref{eq:curie_constant}) are shown as blue dotted lines at low temperatures in Fig. 2, displaying the $1/T$ behavior in the cases $N=4j+1$, $4j+3$, and $4j+4$.

\subsection{High temperature limit}
We observed in Fig. 2 that for all cases of $N$ the susceptibility decays rapidly with increasing temperature. When spin is not included, the susceptibility decays exponentially and experiences an infinite number of sign flips. When spin is included, the susceptibility decays more slowly and either remains paramagnetic or experiences a diamagnetic-paramagnetic transition before approaching zero. This important finding predicts that all rings will display a paramagnetic response at sufficiently high temperatures when spin effects are present. To see why this is the case, we proceed with an analytical evaluation of Eq.~(\ref{eq:susceptibility_sum}) at high temperatures. When the thermal energy is much greater than the energy-level spacing at the Fermi surface ($k_{B}T\gg \epsilon _{F}/N$) or equivalently, $T\gg T_{F}/N$, the energy levels form a nearly continuous band, and the summations in Eqs. (\ref{eq:electron_number_sum}) and (\ref{eq:susceptibility_sum}) can be approximated using integration by applying the Euler-Maclaurin formula (see Appendix for details). The resulting integrals can be evaluated in closed-form using special functions, finding for the chemical potential
\begin{equation}
\label{eq:high_temp_mu}
\mu\left(0\right)=\frac{\epsilon_{F}T}{T_{F}}\mathrm{Log} \left(-\mathrm{Li}_{\frac{1}{2}}^{-1}\left(-2\sqrt{\frac{T_{F}}{\pi T}}\right)\right)
\end{equation}
and the susceptibility
\begin{equation}
\label{eq:high_temp_chi}
\chi =-\frac{4g_{s}^{2}\left| \chi _{L}\right| }{N^{2}}\sqrt{\frac{\pi T_{F}}{T}} \mathrm{Li}_{-\frac{1}{2}}\left(\mathrm{Li}_{\frac{1}{2}}^{-1}\left(-2\sqrt{\frac{T_{F}}{\pi T}}\right)\right)
\end{equation}
where $\mathrm{Li}_{n}(x)$ are the polylogarithm functions of order $n$ and $\mathrm{Li}_{n}^{-1}(x)$ denotes the inverse of the polylogarithm. Making use of the asymptotic values $\mathrm{Li}_{\frac{1}{2}}^{-1}\left(-2\sqrt{T_{F}/(\pi T)}\right)\approx -e^{T_{F}/T}$ and $\mathrm{Li}_{-\frac{1}{2}}\left(-e^{T_{F}/T}\right)\approx -\sqrt{T/(\pi T_{F})} $  in the limit $T\ll T_{F}$, we find from Eq.~(\ref{eq:high_temp_chi})
\begin{equation}
\chi=\frac{4g_{s}^{2}\left| \chi _{L}\right| }{N^{2}}, \qquad \frac{T_{F}}{N}\ll T\ll T_{F}\mathrm{.}
\end{equation}
Thus at high temperatures with spin effects included ($g_{S}\neq 0$), a paramagnetic susceptibility is always expected regardless of $N\,\mathrm{mod}\,4$ (see blue dotted lines on the right sides of Fig. 2). Note that this high temperature limit also serves as the “bulk limit” in Fig. 1 where the criterion $T\gg T_{F}/N$ is satisfied for large $N$. This paramagnetism in the bulk and high temperature limits is purely a spin effect since the magnitude of the susceptibility decays to zero exponentially with increasing temperature when spin is absent.

\begin{figure*}
\includegraphics{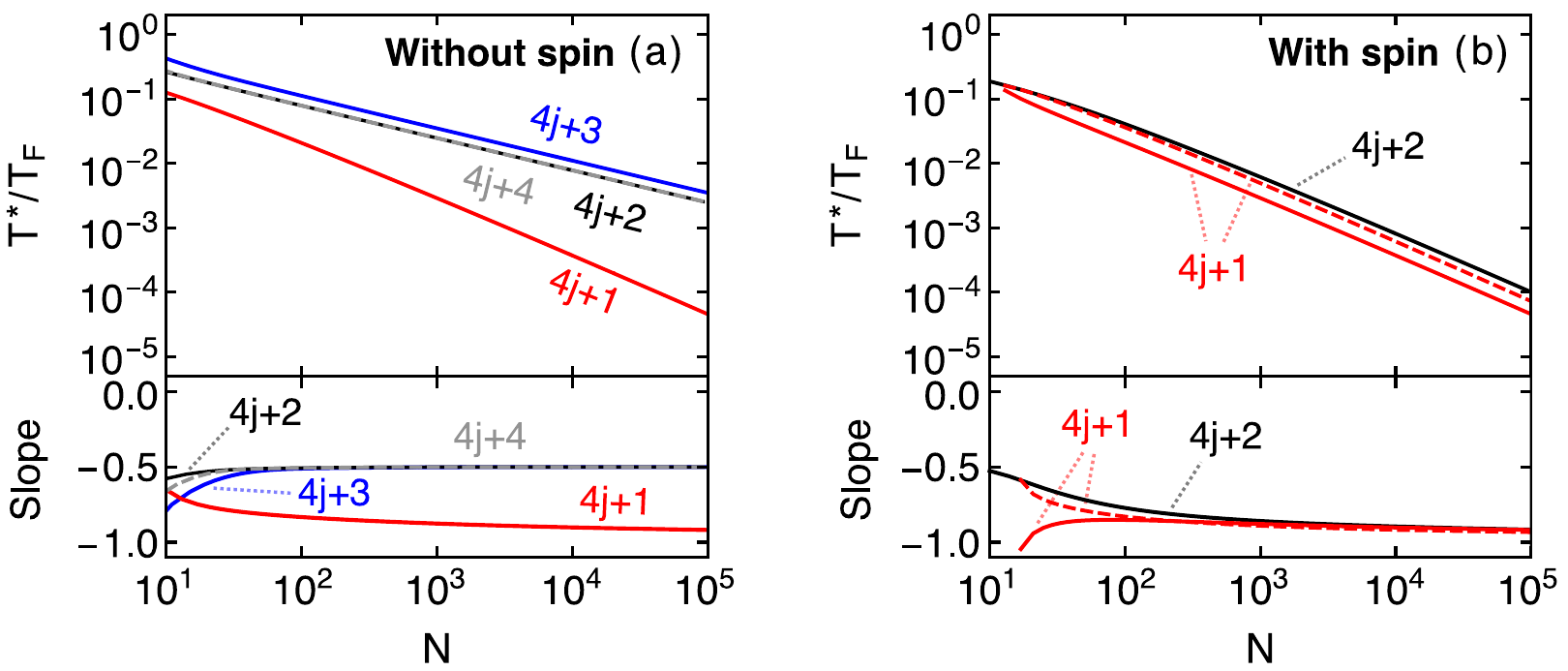}
\caption{
\label{fig:T_crit}
The size dependence of the flipping temperature for each case of $N\,\mathrm{mod}\,4$ when (a) spin effects are turned off ($g_{s}=0$) and (b) spin effects turned on ($g_{s}=2$). The dashed line in (b) is the second flip for the $N=4j+1$ case (see text). The attached bottom panels indicate the slope $p\left(N\right)$ of the respective lines in the top panels, corresponding to the power law $T^{*}/T_{F}\sim N^{p\left(N\right)}$.
}
\end{figure*}

\subsection{Paramagnetic-diamagnetic transitions}
We have already seen in Fig. 2 that the sign of the susceptibility can experience single or multiple flips at intermediate temperature values when $T\approx T_{F}/N$. Figure 3 further visualizes how the transitions between diamagnetism and paramagnetism depend on the number of particles, the temperature, and whether or not spin is included. The diamagnetic H\"{u}ckel rings with $N=4j+2$ are clearly visible as dark blue bars at the bottom of the figure where $T\ll T_{F}/N$. The other three cases for $N$ show strong paramagnetism at low temperatures, shown as dark red bars. As the temperature increases, the susceptibility exponentially decays as it approaches zero in all cases. However, the result of including spin effects becomes apparent with comparing the left and right panels of Fig. 3. From the left panel we see that if only angular momentum plays a role (no spin effects) the susceptibility experiences rapid change of sign as the temperature is increased. On the other hand, in the right panel we observe that when spin effects are included and the ring’s number of electrons corresponds to H\"{u}ckel’s rule ($N=4j+2$) the susceptibility flips only once from diamagnetic to paramagnetic. For rings with $N=4j+1$ the susceptibility flips exactly twice (for $N\geq 13)$ while for all other cases it remains paramagnetic. Furthermore, in accordance to our high temperature analysis (see Eq.~(\ref{eq:high_temp_chi}) and discussions thereafter), all cases manifest a paramagnetic response for sufficiently high temperatures.

Inspecting Figs. 2 and 3, there is clearly a size-dependent transition temperature beyond which paramagnetic-diamagnetic transitions are manifested. We track the size-dependence of this transition by defining the temperature $T^{*}$ to be the lowest temperature at which the susceptibility changes sign. This transition temperature is presented in Fig. 4 where calculations have been performed for each case of $N\,\mathrm{mod}\,4$ with $g_{s}=0$ in Fig. 4(a) and $g_{s}=2$ in Fig. 4(b). The $N=4j+3$ and $N=4j+4$ rings are not shown in Fig. 4(b) since they never change sign when $g_{s}=2$. The dashed line in Fig. 4(b) indicates the second flip of $N=4j+1$ rings so the region between the dashed and solid red lines indicates the range of temperatures in which $N=4j+1$ rings exhibit diamagnetism.

\begin{figure*}
\includegraphics{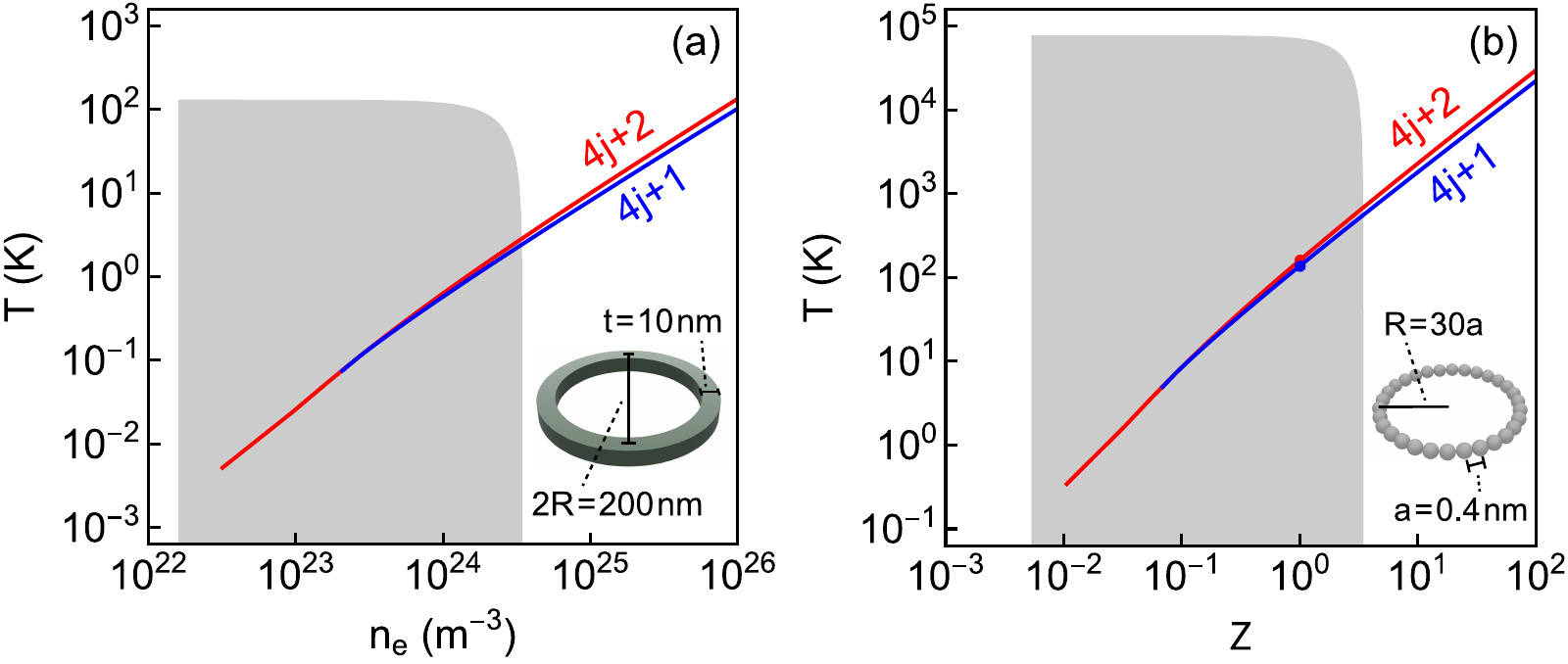}
\caption{
\label{fig:applicability}
The transition temperature calculated for physical rings with $N=4j+1$ or $N=4j+2$ and spin included, modeled after (a) semi-conductor rings with varying electron density and (b) cyclic atom chains with varying valence electron contribution. The shaded areas indicate the regions of applicability of the 1D model ($\Delta \gg k_B T$). The dots in (b) correspond to the contribution of a single electron per atom ($Z = 1$), typical for metals like silver.
}
\end{figure*}

We find that $T^{*}$ decreases rapidly as the number of electrons increases. In particular, the logarithmic plot shows that this temperature follows a size-dependent power law $T^{*}\sim N^{p\left(N\right)}$, where the exponent $p\left(N\right)$, shown in the bottom panels of Fig. 4, corresponds to the slope of the lines in the top panels. For all cases we have $-1\leq p\left(N\right)\leq -1/2$. Denoting by $p_{\infty }$ the asymptotic behavior in the “bulk limit” $N\rightarrow \infty $, in the case when spin effects are not included we obtain from numerical convergence the values of the exponents as
\begin{equation}
	p_{\infty }=\begin{cases} -1, & N=4j+1\\
-1/2, & \text{else}
\end{cases},\qquad g_{s}=0\mathrm{.}
\end{equation}
If spin is included, we obtain
\begin{equation}
	p_{\infty }=\begin{cases} -1, & N=4j+1\\
-1, & N=4j+2
\end{cases},\qquad g_{s}=2\mathrm{.}
\end{equation}
Note that the difference in $p_{\infty }$ between the cases with and without spin can drastically change the temperature at which diamagnetic-paramagnetic transitions appear. This strong discrepancy could potentially serve as an experimental test for the presence of spin effects, a perhaps unexpected result since Zeeman splitting effects are commonly assumed too weak at the small magnetic fields of interest [24]. Experimental verification of the diamagnetic-paramagnetic  transitions would require a detailed temperature study for isolated individual rings of varying size. Modern experiments have already detected large paramagnetic $1/T$ Curie-like temperature dependence in isolated gold rings from $T=$ 25 mK to 0.6 K with an estimated $N\approx 10^{10}$ electrons \cite{Bluhm:2009fe}. However, the predicted temperature-driven transitions between diamagnetism and paramagnetism is yet to be experimentally demonstrated.

\section{Applicability to Physical Rings}
In the analysis above we follow the common practice of modeling rings using a strictly one-dimensional picture \cite{Loss:1991wy}, which applies to thin physical rings with only one transverse mode occupied, \textit{i.e.} the energy gap between the Fermi state and the first excited transverse state is larger than the thermal fluctuations ($\Delta \gg k_B T$). In order to understand how this requirement places a constraint on the material parameters and dimensions of a physical system, we consider a ring with thickness $t$ and $R\gg t$, in which case the energy levels are given by
\begin{equation}
	\epsilon_{m,n,l} = \frac{\hbar^2}{2m_e}\left( \frac{m^2}{R^2} + \frac{\pi^2}{t^2}\left(n^2 + l^2\right) \right)
\end{equation}
where $m$ is the angular quantum number as before, and $n=1,2,...$ and $l=1,2,...$ are the quantum numbers for the transverse directions. The energy gap between the first sub-band and the first excited transverse state can be written
\begin{equation}
\label{eq:energy_gap}
	\Delta = \epsilon_{0,2,1} - \epsilon_{m_F,n,l} = \frac{\hbar^2}{2m_e}\left( \frac{3\pi^2}{t^2} - \frac{m_F^2}{R^2} \right)
\end{equation}
where $m_F$ is the angular quantum number of the highest occupied state. In the following subsections, we consider the applicability condition $\Delta \gg k_B T$
for two different physical configurations of nanosized rings. First, we consider semi-conductor rings with an electron density that can be controlled by varying the dopant concentration. These types of rings have been fabricated using a variety of methods like self-assembly of nanostructures \cite{Lorke:2000tq}, localized oxidation using an atomic force microscope \cite{Fuhrer:2001to}, and nano-lithography techniques \cite{Bayer:2003fm}. The second type of ring we consider is a circular chain of atoms like aromatic carbon molecules and other cyclic compounds \cite{Rickhaus:2020dr} or other ring structures manipulated at the atom-level \cite{VanDongPham:2019jh}.
\vspace{-4mm}
\subsection{Semi-Conductor Rings}
Physical semi-conductor rings are modeled using $m_F = N/4 = n_e \pi R t^2 / 2$ in Eq. (\ref{eq:energy_gap}). Then by solving $\Delta \ll k_B T$ for temperature, we find a condition for applicability of the 1D model,
\begin{equation}
\label{eq:applicabilityA}
	T \ll T_{\mathrm{max}} = \frac{3 \pi^2 \hbar^2}{2 m_e k_B t^2}\left(1 - \left( \frac{n_e t^3}{2\sqrt{3}} \right)^2 \right)
\end{equation}
Additionally, we require at least one electron on the ring, so we must have $n_e \geq 1/(2 \pi R t^2)$ and $t > 2 a_B \approx 0.1$ nm.

To observe a paramagnetic-diamagnetic transition, we must have $T=T^{*}=\hbar^2 \pi^2 n_e^2 t^4 f(N) / (8 m_e k_B) $ where $f(N)$ is the relationship for $T^{*}/T_F$ shown in Fig. \ref{fig:T_crit}. The transition temperature for a quantum ring with dimensions $R = 100$ nm and $t = 10$ nm is shown in Fig. \ref{fig:applicability}(a) for a range of electron densities, spanning from $T^* \approx 25$ mK for $n_e = 10^{23}$ $\mathrm{m}^{-3}$ to $T^* \approx 5$ K for $n_e = 2 \times 10^{24}$ $\mathrm{m}^{-3}$. Because these values fall squarely within the range of applicability for the 1D model given by Eq. (\ref{eq:applicabilityA}) and indicated by the shaded region in Fig. \ref{fig:applicability}, the paramagnetic-diamagnetic transitions should be subject to experimental observation.

\subsection{Atomic Chains}
For chains of atoms, we have $m_F = N/4=Z \pi R / (2 a)$ where $a$ is the lattice constant, and $Z$ is the average number of electrons donated by a single atom to the conduction band. The 1D applicability condition again follows from $\Delta \gg k_B T$ and Eq. (\ref{eq:energy_gap}),
\begin{equation}
	\Delta = \frac{3\pi^2\hbar^2}{2 m_e a^2}\left( 1 - \frac{Z^2}{12} \right) \gg k_B T
\end{equation}
Solving for temperature, the condition becomes
\begin{equation}
	T \ll T_{\mathrm{max}} = \frac{3\pi^2\hbar^2}{2 m_e k_B a^2}\left( 1 - \frac{Z^2}{12} \right)
\end{equation}
Additionally, we must have $N = 2 Z \pi R / a \geq 1$ and $a > a_B$. The transition temperature for atomic chains is then $T^{*}=\hbar^2 \pi^2 n_e^2 Z^2 f(N) / (8 m_e k_B a^2)$. The applicability condition and critical temperature for an atomic chain of silver atoms is shown in Fig. \ref{fig:applicability}(b) with $a = 0.41$ nm and $R=30a$, and the value $T^* \approx 160$ K for $Z = 1$ indicated by the red dot falls entirely within the applicability region. This transition temperature is considerably higher than that of the semi-conducting ring in Fig. \ref{fig:applicability}(b), suggesting that atom chains may be better suited for observation of paramagnetic-diamagnetic transitions approaching room temperatures.

\section{\label{sec:conclusion}Conclusion}
We have investigated the temperature dependence of the magnetic susceptibility in one-dimensional conductive rings for a wide range of temperatures and sizes. Analytical results were provided for the low and high temperature limits. At low temperatures $T\ll T_{F}/N$, conducting rings with a total number of conduction electrons corresponding to H\"{u}ckel’s rule ($N=4j+2$) are always diamagnetic, reaching a maximum value $\chi _{\max }\approx -\chi _{1}T_{F}/T$. For non-H\"{u}ckel rings, a Curie-type paramagnetism is observed. At high temperatures $T\gg T_{F}/N$ and for all values of $N$, the susceptibility exponentially decays with increasing temperature, approaching $\chi \rightarrow 0$ when spin effects are absent. When spin is included, the susceptibility becomes weakly paramagnetic at high temperatures, acquiring the value $\chi =4g_{s}^{2}/N^{2}$ for $T_{F}/N\ll T\ll T_{F}$. For intermediate temperatures higher than a critical value $T^{*}$, our studies show that the conductive ring susceptibility experiences complex behavior including single or multiple diamagnetic-paramagnetic transitions. The bulk limiting behavior of this critical temperature is either $T^{*}\sim N^{-1/2}$ or $T^{*}\sim N^{-1}$ depending on $N\,\mathrm{mod}\,4$ and the presence of spin effects. Finally, the applicability of the 1D model was evaluated for physical semiconductor rings and cyclic atom chains, showing that the predicted magnetic transitions should be subject to experimental studies.
\appendix\label{appendix:sum_rule}
\section{Chemical potential in the high temperature limit}
In this appendix, we provide details of the calculation of the high-temperature result for the chemical potential. Written explicitly using Eqs. (\ref{eq:electron_number_sum}) and (\ref{eq:eigenenergies}), the chemical potential is defined by
\begin{align*}
	N=\sum _{m=-\infty }^{\infty }\left(\frac{1}{1+e^{\frac{\epsilon_1 }{k_{B}T}\left(\left(m+\kappa \right)^{2}+g_{s}\kappa -\tilde{\mu }\right)}} \right. \nonumber \\ + \left. \frac{1}{1+e^{\frac{\epsilon_1 }{k_{B}T}\left(\left(m+\kappa \right)^{2}-g_{s}\kappa -\tilde{\mu }\right)}}\right)\mathrm{.}
\end{align*}
Changing the limits of the summation, we find the equivalent form
\begin{equation}
\label{eq:a1}
	N=-\frac{f(0)}{2}+\sum _{m=0}^{\infty }f(m)
\end{equation}
where
\begin{align*}
f(m)=&\frac{1}{1+e^{\frac{\epsilon_1 }{k_{B}T}\left(\left(m+\kappa \right)^{2}+g_{s}\kappa -\tilde{\mu }\right)}} \\ &+ \frac{1}{1+e^{\frac{\epsilon_1 }{k_{B}T}\left(\left(m+\kappa \right)^{2}-g_{s}\kappa -\tilde{\mu }\right)}} \\ &+ \frac{1}{1+e^{\frac{\epsilon_1 }{k_{B}T}\left(\left(-m+\kappa \right)^{2}+g_{s}\kappa -\tilde{\mu }\right)}} \\ &+\frac{1}{1+e^{\frac{\epsilon_1 }{k_{B}T}\left(\left(-m+\kappa \right)^{2}-g_{s}\kappa -\tilde{\mu }\right)}}\mathrm{.}
\end{align*}
Applying the Euler-Maclaurin formula \cite{Arfken:0uf} we find $\sum _{m=0}^{\infty }f(m)=\int _{0}^{\infty }f(m)\,dm+f(0)/2$ where we have used that $f^{(2k-1)}(0)=f^{(2k-1)}(\infty)=f(\infty)=0$. With this result and Eq.~(\ref{eq:a1}), we can now write$\newline $
\begin{equation*}
N=\int _{0}^{\infty }f(m)\,dm\mathrm{.}
\end{equation*}
With the substitutions $x=m+\kappa $ and $y=m-\kappa $,  we can split the integral in the following way
\begin{align*}
\int _{0}^{\infty }f(m)\,dm=\int _{0}^{\infty }F(x)\,dx-\int _{0}^{\kappa }F(x)\,dx \\ +\int _{0}^{\infty }G(y)\,dy-\int _{\kappa }^{0}G(y)\,dy
\end{align*}
where the integrands are defined by
\begin{align*}
F(x)&=\frac{1}{1+e^{\frac{\epsilon_1 }{k_{B}T}\left(x^{2}+g_{s}\kappa -\tilde{\mu }\right)}}+\frac{1}{1+e^{\frac{\epsilon_1 }{k_{B}T}\left(x^{2}-g_{s}\kappa -\tilde{\mu }\right)}} \\
G(y)&=\frac{1}{1+e^{\frac{\epsilon_1 }{k_{B}T}\left(y^{2}+g_{s}\kappa -\tilde{\mu }\right)}}+\frac{1}{1+e^{\frac{\epsilon_1 }{k_{B}T}\left(y^{2}-g_{s}\kappa -\tilde{\mu }\right)}} \mathrm{.}
\end{align*}
Recognizing that $x$ and $y$ are dummy integration variables, we have
\begin{align*}
\int _{0}^{\kappa }F(x)\,dx &=-\int _{\kappa }^{0}G(y)\,dy \\ \int _{0}^{\infty }F(x)\,dx &= \int _{0}^{\infty }G(y)\,dy
\end{align*}
so we are left with $\int _{0}^{\infty }f(m)\,dm=2\int _{0}^{\infty }F(x)\,dx$, giving
\begin{align*}
N = 2\int _{0}^{\infty }&\left(\frac{1}{1+e^{\frac{\epsilon_1 }{k_{B}T}\left(x^{2}+g_{s}\kappa -\tilde{\mu }\right)}} \right. \\ & \quad\qquad +  \left. \frac{1}{1+e^{\frac{\epsilon_1 }{k_{B}T}\left(x^{2}-g_{s}\kappa -\tilde{\mu }\right)}}\right)dx\mathrm{.}
\end{align*}
This integral can be evaluated using polylogarithm functions, leading to a transcendental equation for $\tilde{\mu },$
\begin{align}
\begin{aligned}
\label{eq:a2}
	N=-\sqrt{\frac{\pi k_{B}T}{\epsilon_1 }}&\left(\mathrm{Li}_{\frac{1}{2}}\left(-e^{\frac{\epsilon_1 }{k_{B}T}\left(\tilde{\mu }-g_{s}\kappa \right)}\right) \right. \\ 
	 & \quad + \left. \mathrm{Li}_{\frac{1}{2}}\left(-e^{\frac{\epsilon_1 }{k_{B}T}\left(\tilde{\mu }+g_{s}\kappa \right)}\right)\right)
	\end{aligned}
\end{align}
where $\mathrm{Li}_{n}(x)$ are the polylogarithm functions of order $n$. Evaluating Eq.~(\ref{eq:a2}) at $\kappa =0$ gives
\begin{equation*}
	N=-2\sqrt{\frac{\pi k_{B}T}{\epsilon_1 }}\mathrm{Li}_{\frac{1}{2}}\left(-e^{\frac{\epsilon_1 \tilde{\mu }\left(0\right)}{k_{B}T}}\right)\mathrm{,}
\end{equation*}
and substituting $T_{F}$ for $\epsilon_1 $ via Eq.~(\ref{eq:fermi_temp}), we arrive at the result
\begin{equation}
\label{eq:a3}
	-1=\frac{\sqrt{\pi}}{2}\left(\frac{T}{T_{F}}\right)^{1/2}\mathrm{Li}_{\frac{1}{2}}\left(-e^{\frac{T_{F}\mu\left(0\right)}{\epsilon_F T}}\right)
\end{equation}
Finally we find Eq.~(\ref{eq:high_temp_mu}) when we solve Eq.~(\ref{eq:a3}) for the chemical potential. Making use of the asymptotic value $-\mathrm{Li}_{\frac{1}{2}}^{-1}\left(-2\sqrt{T_{F}/(\pi T)}\right)\approx e^{\frac{T_{F}}{T}}$ for $T\ll T_{F}$ recovers the “bulk” result $\mu\left(0\right)=\epsilon_F=\epsilon_1 N^{2}/16$ quoted in the text.
\section{Susceptibility in the high temperature limit}
Here we derive the high-temperature susceptibility result given by Eq.~(\ref{eq:high_temp_chi}) in the text. We begin by changing the limits of the summation in Eq.~(\ref{eq:susceptibility_sum}), finding the equivalent form
\begin{align*}
\chi &= \left| \chi _{L}\right| \left(-1+\frac{2b}{N}\sum _{m=-\infty }^{\infty }f(m)\right) \\ &= \left| \chi _{L}\right| \left(-1+\frac{4b}{N}\left(-\frac{f(0)}{2}+\sum _{m=0}^{\infty }f(m)\right)\right)
\end{align*}
where $b=8T_{F}/(N^{2}T)$ and
\begin{equation*}
f(m)=\left(m^{2}+\frac{g_{s}^{2}}{4}\right)\sech^{2} \left(b\left(m^{2}-\tilde{\mu }(0)\right)\right)
\end{equation*}
Now applying the Euler-Maclaurin formula in the same way as we did for the chemical potential in Appendix A, we find $\sum _{m=0}^{\infty }f(m)=\int _{0}^{\infty }f(m)\,dm+f(0)/2$ since again we have $f^{(2k-1)}(0)=f^{(2k-1)}(\infty)=f(\infty)$. This leads to
\begin{equation}
\label{eq:b1}
	\chi=\left| \chi _{L}\right| \left(-1+\frac{4b}{N}\int _{0}^{\infty }f(m)\,dm\right)\mathrm{,}
\end{equation}
and with the substitution $x=bm^{2}$ the integral becomes
\begin{align*}
\int _{0}^{\infty }f(m)dm=&\frac{1}{2b^{3/2}} \int _{0}^{\infty }x^{1/2}\,\sech^{2} \left(x-b\tilde{\mu}(0)\right)dx \\ &+\frac{g_{s}^2}{8b^{1/2}}\int _{0}^{\infty}x^{-1/2}\,\sech^{2} \left(x-b\tilde{\mu}(0)\right)dx\mathrm{.}
\end{align*}
With the help of the integral formula
\begin{equation*}
\int _{0}^{\infty }x^{n}\,\sech^{2} \left(x-a\right)dx=-2^{1-n}\,\Gamma\left(n+1\right)\mathrm{Li}_{n}\left(-e ^{2a}\right)
\end{equation*}
valid for $n>-1$, we find
\begin{align*}
	\int _{0}^{\infty }f(m)\,dm =& -\frac{1}{2}\sqrt{\frac{\pi}{2}}\left(\frac{1}{b^{3/2}}\mathrm{Li}_{\frac{1}{2}}\left(-e ^{2 b \tilde{\mu}(0)}\right) \right. \\ & \qquad\qquad + \left. \frac{g_{s}^{2}}{b^{1/2}}\,\mathrm{Li}_{-\frac{1}{2}}\left(-e ^{2 b \tilde{\mu}(0)}\right)\right)\mathrm{.}
\end{align*}
With this result, Eq.~(\ref{eq:b1}) becomes
\begin{align}
	\begin{aligned}
	\label{eq:b2}
	\chi =& \left| \chi _{L}\right| \left(-1-\frac{\sqrt{\pi }}{2}\left(\frac{T}{T_{F}}\right)^{1/2}\mathrm{Li}_{\frac{1}{2}}\left(-e^{\frac{\mathrm{T}_{\mathrm{F}}\mu(0)}{\epsilon_{F}T}}\right) \right. \\ 
	& \left. -\frac{4g_{s}^{2}\sqrt{\pi}}{N^{2}}\left(\frac{T}{T_{F}}\right)^{-1/2}\,\mathrm{Li}_{-\frac{1}{2}}\left(-e^{\frac{\mathrm{T}_{\mathrm{F}}\mu(0)}{\epsilon_{F}T}}\right)\right)\mathrm{.}
	\end{aligned}
\end{align}
Inserting the relationship Eq.~(\ref{eq:a3}) into Eq.~(\ref{eq:b2}), we find
\begin{align}
\label{eq:b3}
	\hspace{-1em}\chi=-\frac{4g_{s}^{2}\sqrt{\pi}\left| \chi _{L}\right|}{N^{2}}\left(\frac{T}{T_{F}}\right)^{-\frac{1}{2}}\,\mathrm{Li}_{-\frac{1}{2}}\left(-e ^{\frac{\mathrm{T}_{\mathrm{F}}\mu\left(0\right)}{\epsilon_F T}}\right)\mathrm{.}
\end{align}
Finally, inserting Eq.~(\ref{eq:high_temp_mu}) into Eq.~(\ref{eq:b3}) recovers Eq.~(\ref{eq:high_temp_chi}) given in the text.

\begin{acknowledgments}
This work was supported by the NSF EPSCoR CIMM project under award \#OIA-1541079 and the Louisiana Board of Regents.
\end{acknowledgments}

\bibliography{ref}

\end{document}